\documentclass[12pt]{iopart}

\usepackage{epsfig}
\begin{document}

\title[Cosmic Ray Astronomy]{Cosmic Ray Astronomy}

\author{P Sommers}

\address{Center for Particle Astrophysics, Physics Department, 
Penn State, University Park, PA 16802, USA}
\ead{sommers@phys.psu.edu}

\author{S Westerhoff}
\address{Department of Physics, University of Wisconsin,
Madison, WI 53706, USA}
\ead{westerhoff@physics.wisc.edu}

\begin{abstract}

Cosmic ray astronomy attempts to identify and study the sources
of ultrahigh energy cosmic rays.  It is unique in its reliance on {\it
charged} particles as the information carriers.  While no discrete source of
ultrahigh energy cosmic rays has been identified so far, a new
generation of detectors is acquiring the huge exposure that is needed
at the highest energies, where deflection by magnetic fields is
minimized and the background from distant sources is eliminated by
pion photoproduction.  In this paper, we summarize the status of
cosmic ray astronomy, describing the detectors and the analysis
techniques. 

\end{abstract}

\maketitle

\section{Introduction}

The goal of cosmic ray observatories is to discover the sources
of the Universe's highest energy particles, then to use cosmic rays
to study the properties of these sources and of the magnetic fields that
the cosmic rays have traversed.  Individual cosmic ray particles have 
been observed to have energies as great as 50 Joules, and the mystery 
of how they are produced has persisted for 45 years~\cite{linsley:1963}.
Evidence now indicates that they are accelerated by electromagnetic 
processes, not produced by the decay of supermassive particles
(see Sec.~\ref{subsec:gamma}).  While the energy spectrum and chemical
composition of the cosmic rays are potential clues to the
astrophysical accelerators, positive identification of the sources can
only be achieved by using anisotropy of their arrival directions to
reveal where the sources are.  

Cosmic ray astronomy differs from electromagnetic, neutrino, and
gravitational wave astronomy due to the electric charge of the
messengers and is sometimes called charged particle astronomy.
All other astronomy relies on neutral particle messengers.

Although there is no guarantee that discrete sources will be
observable with cosmic rays, there are certainly reasons to be
optimistic.  The Larmor radius $R_L$ for a particle of energy $E$ and
charge $Z$ in a magnetic field $B$ is roughly
\begin{equation}
R_L\simeq 1\,\mathrm{kpc}\frac{E\left[\mathrm{EeV}\right]}
{Z\,B\left[\mathrm{\mu G}\right]}~~.
\end{equation}
For cosmic ray particles above several EeV (1\,EeV=$10^{18}$\,eV), the 
Larmor radius exceeds galactic dimensions for typical magnetic fields 
of $\mathrm{\mu G}$ strength.  The highest energy cosmic rays could 
have sufficient magnetic rigidity to maintain their direction to better 
than a few degrees while traveling through the Galaxy, assuming they 
are protons.  With sufficient exposure, the arrival directions from a 
strong nearby source should stand out as a high-density cluster on the
sky.  Moreover, above approximately $6\times 10^{19}$\,eV, there should 
be essentially no background, because the cosmic background radiation
removes such energetic particles before they can travel as much as 
100\,Mpc.  Protons lose their energy to pion photoproduction; nuclei
photodisintegrate; and gamma rays succumb to $e^{\pm}$ pair creation.
With the isotropic background from more distant parts of the Universe
eliminated, only the foreground sources contribute.

The onset of pion photoproduction from the interaction of protons with the
microwave background radiation at $6\times 10^{19}$\,eV is expected
to lead to a suppression of the cosmic ray flux above this threshold
energy.  Observational evidence for this so-called Greisen-Zatsepin-Kuz'min 
(GZK) suppression~\cite{greisen:1966,zatsepin:1966} has been mounting 
in recent years after the Akeno Giant Air Shower Array
(AGASA), with the largest cumulative exposure at that time, cast doubt
upon it~\cite{takeda:1998,takeda:2003}.  The High Resolution Fly's Eye
experiment~\cite{abbasi:2007} has claimed statistically significant
evidence for the expected suppression of a power-law spectrum.  The
most recent spectrum from data taken with the Pierre Auger Observatory 
in Argentina is also consistent with the GZK
suppression~\cite{yamamoto:2007}.  It appears that cosmic rays above
the GZK energy threshold must originate in the nearby Universe, with
source distances not exceeding 100\,Mpc.  Searches for the sources of
cosmic rays can therefore expect a background-free search for
astrophysical objects within this ``GZK-sphere'' by selecting events
above the GZK threshold energy.  These super-GZK cosmic rays are at least a
million times more energetic than any other messengers that have been
detected from astrophysical sources.

The reward of measuring the cosmic rays from an individual source will
not be limited to new knowledge about the astrophysical source itself.
By studying how the arrival directions deviate from the source
direction as a function of particle energy, it should be possible to
infer valuable information about intervening magnetic fields that
exist between the Earth and the source (see for 
example~\cite{stanev:1997,medina:1998,prouza:2003}).

The fact that no prominent source has so far been detected casts some
doubt on this optimistic scenario.  If there are indeed astrophysical
sources of cosmic rays that can be identified by arrival directions,
it is likely that the sources will first be detected collectively
rather than individually.  A class of sources can be identified
statistically even if no single source is detectable.

Autocorrelation studies, for example, may reveal an excess of pairs 
and triplets of clustered arrival directions.  Although any of the 
multiplets could occur by chance, an implausibly large number of them 
would constitute compelling evidence of discrete sources.  But even 
without any multiplets of arrival directions, the sources of cosmic 
rays can be identified by a correlation of the (singlet) arrival 
directions with a class of candidate astrophysical objects.  If the 
observed correlation were extremely unlikely to occur by chance
from an isotropic cosmic ray intensity, then the correlation could be
regarded as a special anisotropy consistent with the hypothesis that
the class of astrophysical objects include sources of cosmic rays.
This can happen even if the cosmic ray arrival directions show no
clustering or large-scale pattern.  Finding compelling autocorrelation 
of arrival directions or correlation of arrival directions with a class 
of astrophysical objects would validate the assumption that discrete
sources exist and that the sources can be studied individually
when sufficient cosmic ray exposure is obtained.  From the Pierre Auger 
Observatory, there is now indeed evidence for a correlation between 
the arrival directions of ultrahigh energy cosmic rays and the positions 
of nearby Active Galactic Nuclei (AGN)~\cite{abraham:2007c,abraham:2007d}.   
These results, described in Sec.~\ref{sec:auger_agn}, raise prospects
of finally identifying individual sources with more data in the near future.

Information about the origins of cosmic rays should also be
extractable from large-scale patterns in the celestial distribution of
arrival directions.  A cosmic ray gradient can be detected as a
dipole.  A quadrupole pattern may be expected if sources lie near some
plane in space.  Large-scale patterns may be the best handle on cosmic
ray origins if they are not produced by discrete sources or if the
discrete sources cannot be resolved.  Large-scale patterns can be
expected to be effective in the study of galactic production of cosmic
rays near 1 EeV.  Large observatories will obtain sky maps with high
statistics in that energy regime.  In order to exploit the power of
spherical harmonics for characterizing the anisotropy, it is important
that observatories acquire good exposure to the entire celestial
sphere by having sites in both the northern and southern hemispheres.

The search for discrete cosmic ray sources occurs at the highest
energies, where magnetic deflections should be small.  The challenge
of cosmic ray astronomy is the fact that the flux of particles is tiny
at such high energy, so enormous exposure is needed.  Of course a major
uncertainty is the strength of the galactic and intergalactic magnetic
fields.  Estimates of the field strength vary widely.  In~\cite{dolag:2004},
Dolag {\it et al.} model intergalactic magnetic fields with
magneto-hydrodynamical simulation of cosmic structure formation, using
the (few) existing measurements based on Faraday rotation near
clusters as boundary conditions.  The authors find that over large
parts of the sky, the angular deflection for protons of energy
$4\times 10^{19}$\,eV should be of order $1^{\circ}$ for propagation
distances up to 500\,Mpc, with larger deflections occurring in cluster
regions.  However, Sigl {\it et al.} estimate that deflections might
be much larger~\cite{sigl:2004}.

The galactic magnetic field is better known, but here, too, large 
uncertainties remain (see~\cite{widrow:2002} for a review of experimental
methods and results).  The galactic magnetic field has a regular and a random
(turbulent) component which affect the trajectories of charged particles
in different ways.  Estimates of the strength of the {\it regular} component
exist from optical and synchrotron polarization measurements~\cite{beck:2001} 
and Faraday rotation measures of pulsars~\cite{han:2006}; with their different 
systematics, polarization measurements tend to overestimate the field strength, 
whereas rotation measures tend to underestimate it.  The experimental results
indicate that the magnitude of the regular field in our local neighborhood is 
of order 2 to 4\,$\mu$G.  There is a lot of uncertainty on how the field falls
off along the direction perpendicular to the disk and in the galactic halo,
which will strongly affect cosmic ray trajectories.  As a rough estimate, a
cosmic ray proton of energy $6\times 10^{19}$\,eV traveling through a
regular galactic field of strength 3\,$\mu$G for 1\,kpc 
is deflected by $3^{\circ}$.  Simulations based on more extensive models of 
the structure of the field, including the field in the 
galactic halo, essentially give values for the deflection consistent
with this rough estimate~\cite{stanev:1997,medina:1998,prouza:2003}.
Being proportional to the charge of the cosmic ray particle, deflections 
are much larger for heavier nuclei.

Deflections in the {\it random} component of the galactic magnetic field have
been estimated in~\cite{tinyakov:2005}.  While the magnitude of the field 
strengths of the random and the regular component are comparable, deflections 
by the random component are expected to be smaller, since the correlation 
length of the random fields is only about 50 to 150\,pc.  A path of 1\,kpc 
length would experience multiple small deflections, with no systematic change 
in direction.  The mean square deflection due to the random fields is estimated 
in~\cite{tinyakov:2005} to be smaller than deflections due to the regular 
component by a factor ranging between 0.03 and 0.3.

There is substantial quantitative uncertainty about the magnetic smearing 
of point sources due to regular and random magnetic fields within the Galaxy 
as well as in intergalactic space.  For energies near and above the GZK 
threshold, however, it is certainly possible that protons arriving from a 
point source could be concentrated within a circle less than a few degrees 
in radius.  While not guaranteed, charged particle astronomy is a realistic 
possibility.

Discovering the sources of cosmic rays (at all energies) has been a
motivating objective for very high energy gamma ray observatories and
also for neutrino detectors.  Gamma ray astronomy now studies astrophysical
objects in gamma rays with energies up to tens of TeV.  These are
currently the highest energy particles whose sources have been
unambiguously identified.  A number of galactic and extragalactic
sources have been identified, and new sources are now discovered on a
regular basis with telescopes in both hemispheres (see the article
by J. Hinton in  this Focus Issue).
Above several tens of TeV, gamma ray astronomy faces a limit, as
interactions with extragalactic photon fields, mainly the infrared and
microwave backgrounds, severely restrict the distances over which
gamma rays can travel unattenuated by $\gamma \gamma \rightarrow
e^+ e^-$ pair production.

In many ways, neutrinos are the ideal messenger particles for astronomy
at the highest energies.  Like gamma rays, neutrinos travel undeflected
by magnetic fields, but unlike gamma rays, they are not absorbed in radiation 
fields.  Of course the small neutrino cross section, while ideal
to reveal possible sources, has a downside.  Neutrinos are not easily
absorbed in detectors either, and the major challenge of neutrino
astronomy is to build detectors that can achieve a decent detection rate of
astrophysical neutrinos.  That requires at least kilometer-scale detectors.
No sources of astrophysical neutrinos have been identified
so far except the supernova explosion SN1987A and the Sun, both 
detected via neutrinos in the MeV range.  

In this Paper, we summarize the prospects of charged particle astronomy. 
After a section on the conditions for discrete source detection
(Sec.~\ref{sec:conditions}), we will discuss cosmic ray detection techniques
and describe several past, current, and planned experiments 
(Sec.~\ref{sec:experiments}).  Section~\ref{sec:claims} reviews past claims 
for discoveries of cosmic ray sources and describes some of the analysis 
techniques applied in searches for sources.  In Section~\ref{sec:neutrino},
we review the current status of astronomy with ultrahigh energy 
gamma rays and neutrinos.  A section on the importance of full-sky coverage
for the future of cosmic ray astronomy (Sec.~\ref{sec:fullsky}) and a summary 
(Sec.~\ref{sec:summary}) conclude the paper.

\section{Conditions for discrete source detection}
\label{sec:conditions}

From the measured cosmic ray spectrum, we know that the intensity of
cosmic rays is minuscule (of order several $10^{-3}\,\mathrm{km}^{-2}\,
\mathrm{sr}^{-1}\,\mathrm{yr}^{-1}$) at and above the GZK threshold.  
Enormous exposure is necessary to acquire enough arrival directions
to obtain statistically compelling evidence for individual discrete
sources.  The fact that none has been discovered so far means that a
big increase in exposure will be needed.  An exposure of 
$10^{5}\,\mathrm{km}^{2}\,\mathrm{sr}\,\mathrm{yr}$ would yield several 
hundred arrival directions, roughly an order of magnitude more than has 
been obtained with the greatest exposure so far.
Still, there is no guarantee that cosmic rays are accelerated at isolated 
discrete sources.  Even if they are, the sources may be too numerous and
individually weak.  They may be transient or burst sources.  They may
emit only in narrow jets.  

Here we will examine the dependence on the mean separation $R$ between
sources, assuming all of the sources to be identical steady sources
which emit cosmic rays in all directions with a luminosity $Q$ (cosmic
rays per unit time).  It is straightforward to show that the nearest
source is more detectable if the mean separation $R$ is large.  For a
high density of sources (small $R$), detection is difficult because
each source is then very weak.

The measured energy spectrum tells us the intensity $I(>E)$ above any
threshold energy.  (This is the number of cosmic rays per unit area
per unit solid angle per unit time.)  Knowing the intensity is
equivalent to knowing the density of cosmic rays
$\rho=\frac{4\pi}{c}I$, where $c$ is the speed of light and $I$ stands
for the integral intensity $I(>E)$ above the energy threshold.  This
spatial density of cosmic rays must be given by $Q$ times the density
of sources ($1/R^3$) times the time of accumulation $T$.  For
super-GZK particles with an attenuation range of 100\,Mpc, for example, 
the accumulation time would be $100\,\mathrm{Mpc}/c$, whereas it
would be at least an order of magnitude greater below the GZK
threshold.  Equating the two expressions for the cosmic ray density
gives the luminosity $Q$ of each source: $Q=4\pi I R^3/(cT)$.  The
luminosity $Q\sim R^3$ means that the luminosity of each source must
be very high if the mean distance $R$ between sources is large.  For a
source at distance $r$ from the Earth, the flux is $Q/(4\pi r^2)$ or
$I R^3/(cTr^2)$.  (Here it is assumed that the distance r is small
compared to $cT$ so that the flux is not greatly attenuated by the GZK
effect.)  The nearest source to Earth is likely to be at a distance
that is comparable to the mean spacing $R$.  Using $r\sim R$, the flux is
$\sim IR/(cT)$.  That is, the expected flux from the nearest source
increases in proportion to the mean separation $R$.  Discrete sources
are more likely to be detectable if the mean separation $R$ is large
(but not greater than the GZK attenuation length).  Further
analysis has been presented elsewhere \cite{sommers:2007} regarding
the factors that govern the detectability of discrete sources of
cosmic rays both above and below the GZK threshold.

\section{Cosmic ray detectors}
\label{sec:experiments}

The primary cosmic ray flux is a steeply falling power law in energy,
and the integrated flux above $10^{19}$\,eV is only about one particle
per square kilometer per year.  This rules out direct detection of the 
primary cosmic ray particle with
satellite or balloon-borne detectors, as large areas are required
to accumulate reasonable statistics.  Consequently, cosmic ray
detectors at ultrahigh energies use the Earth's
atmosphere as the detector medium.  The primary cosmic ray particles
are not observed directly, since they interact in the upper atmosphere
and induce extensive air showers with roughly $10^{10}$ particles for
a $10^{19}$\,eV primary.  The properties of the original cosmic ray
particle, such as arrival direction and energy, have to be inferred
from the measured properties of the extensive air shower.

Two detector types have traditionally been used to record air showers:
surface detector arrays and air fluorescence detectors.  The former
consist of arrays of particle detectors on the Earth's surface that
sample the particles of the air shower cascade that reach the
ground. The AGASA experiment in Japan is an example of a pure surface
detector array.  In its final configuration, AGASA comprised 110
scintillation counters separated by 1000\,m and spread over an area
of $100\,\mathrm{km}^2$.  The array took data between 1987 and 2003
and collected a total of 72 events with zenith angles smaller than
$45^{\circ}$ above $4\times 10^{19}$\,eV.  It achieved an angular 
resolution of $2.5^{\circ}$ (68\,\%) at energies above
$10^{19}$\,eV.

Scintillation counters are not the only possible basic unit for 
surface detector arrays.  The Haverah Park experiment, a 
$12\,\mathrm{km}^{2}$ air shower array operated in England between 
1967 and 1987~\cite{lawrence:1991}, pioneered the use of water Cherenkov 
detectors.  The particle detectors are light-tight water tanks with 
photomultiplier tubes inside.  The tubes detect the Cherenkov radiation 
from air shower particles crossing the tanks. 

Surface detector arrays sample the air shower at one altitude only 
and do not record the development of the shower in the atmosphere.  
The total energy of the air shower and thus the incident cosmic ray 
primary is connected to the measured particle density on the ground
by air shower simulations, making the energy determination of cosmic 
rays with ground arrays model-dependent.

In contrast, the air fluorescence technique is able to image the shower
development in the atmosphere.  The measured fluorescence light is
produced when particles of the extensive air shower interact with
nitrogen molecules in the atmosphere.  A nearly calorimetric energy
measurement is obtained, because the fluorescence light produced is
proportional to the energy dissipated in the atmosphere.  With the air
fluorescence technique, quantities like the height of the shower's
maximum size can be determined directly.  If the shower is viewed
simultaneously by two detectors in stereo mode, the arrival direction
can be reconstructed with an accuracy of less than $1^{\circ}$.  The
main shortcoming of the technique is the low duty cycle of only about
$10\,\%$, as air fluorescence detectors can only be operated on dark,
moonless nights with good atmospheric conditions, when the small
fluorescence signal can be distinguished from the night sky
background.

An example of a pure air fluorescence detector, the High Resolution
Fly's Eye (HiRes), operated between 1997 and 2003 on the
Dugway Proving Grounds in Utah.  With the completion of a second site,
HiRes operated in stereo mode after 1999 and collected a total of
44 stereo events above $4\times 10^{19}$\,eV, with an
angular resolution of $0.6^{\circ}$ (68\,\%) above $10^{19}$\,eV.

The measurements of cosmic ray air showers by ground arrays and air
fluorescence detectors have fundamentally different systematic errors.
Surface detector arrays depend on the numerical simulations of air
showers, including hadronic interactions that occur at energies well
above experimental verification.  
Air fluorescence measurements of air shower energies, on the other 
hand, are nearly calorimetric, and the largest contribution to the 
systematic error comes from what is presently a 15\,
the proportionality between air fluorescence and charged particle 
energy loss in the atmosphere.  Moreover, atmospheric attenuation of
the fluorescence light on its way to the detectors must be accurately
known.  Scattering of the fluorescence light by aerosol particles
causes a variable amount of attenuation which must be monitored
rigorously.  The aerosols also affect the amount of forward Cherenkov
light which is scattered to the detectors and collected together with
the fluorescence light.  Clouds are more obvious obstructions of
fluorescence light that require continuous automated monitoring.  In
addition, absolute calibration of the fluorescence detectors (photon
flux per analog-to-digital-converter(ADC)) is a challenge since there 
is no naturally occurring calibrating flux.  On the other hand, nature 
does provide an abundance of muons which allow continual calibration 
of the ``vertical equivalent muon'' signal level for a water Cherenkov 
surface detector.  The individual surface detector stations are thereby
automatically calibrated, but the measurement of air shower energy relies 
on air shower simulations.  The fluorescence detector provides the air 
shower energy by a calorimetric method that is simple in principle, but
determining the energy deposition from the detector signal is not at
all simple in practice.  

The two types of detection are complementary
in other respects as well.  The surface array operates continuously
with an acceptance which is uniform over the array and which is
independent of energy (above its full-efficiency threshold).  In
contrast, the fluorescence detector has an irregular duty cycle
governed by weather, sun, and moon.  Its aperture varies with time as
atmospheric conditions change, and the aperture is always larger for
higher energy showers which produce more light and so can be seen from
a larger distance.  On the other hand, the fluorescence detectors
measure showers equally at all zenith angles, whereas the response of
the surface detector depends on the shower inclination.  Because of
their different systematic errors, the two independent detection
techniques are complementary, and the new generation of cosmic ray
experiments uses {\it both} techniques at the same site to 
check the systematic errors in each of them and to obtain a maximum
amount of information for the subset of showers that are recorded
during clear dark nights.

The Pierre Auger Southern Observatory, currently nearing completion in
Malarg\"ue, Argentina, is the first modern {\it hybrid} cosmic ray
experiment, and the first large cosmic ray experiment operating in the
southern hemisphere.  In its final stage, the surface detector array
of the Pierre Auger Southern Observatory will comprise 1600 water
Cherenkov detector tanks, deployed over an area of
$3000\,\mathrm{km}^{2}$ using a regular grid of triangles with 1500\,m
distance between nearest neighbors.  Three photomultiplier tubes in
each of the light-tight tanks measure the water Cherenkov light
produced by the particles of the extensive air shower cascade that hit
the tank.  In addition, four fluorescence detector stations overlook
the surface detector array from the periphery, each with a field of
view of $180^{\circ}$ in azimuth and covering an elevation angle range
from $1.6^{\circ}$ to $30.2^{\circ}$ above the horizon.  Using
``hybrid events,'' {\it i.e.} events seen both in the surface and the
fluorescence detector, the observatory can calibrate the surface
detector array without reference to air shower simulations.  The
calorimetric energy measurement of the fluorescence detector is found
to be proportional to the water tank signal on the ground 1000\,m
from the core.

The Pierre Auger Observatory started scientific data taking in January
2004 and has published first physics results~\cite{abraham:2007c,
abraham:2007d,abraham:2007,abraham:2007b,abraham:2007e,abraham:2007f} .  
With the surface detector alone, it achieves an angular resolution of 
$0.9^{\circ}$ (68\,\%) above $10^{19}$\,eV.  With the fluorescence detector 
and at least one surface detector station, the angular resolution is
about $0.6^{\circ}$.

The ``hybrid'' concept is also used by the Telescope Array (TA) 
experiment~\cite{abbasi:2007b} currently under construction near Delta, 
Utah.  Upon completion, the experiment will consist of 576 double-layer 
scintillation counters with $3\,\mathrm{m}^2$ area each, deployed over an
area of $1000\,\mathrm{km}^{2}$ with 1.2\,km spacing between the counters,
and three fluorescence detectors overlooking the surface detector 
array, each covering $108^{\circ}$ in azimuth and $3^{\circ}-31^{\circ}$ 
in elevation.  For the surface detector array, an angular resolution of 
$1.5^{\circ}$ is expected. 

\section{Prior Hints for Cosmic Ray Sources}
\label{sec:claims}

The small world data set of cosmic ray arrival
directions of the pre-Auger era, consisting of little more than 100
events with energies above $4\times 10^{19}$\,eV, has been subjected to
intense searches for anisotropies on all angular scales, including a
variety of attempts to correlate classes of known astrophysical
objects with cosmic ray arrival directions.

None of the earlier efforts to identify the sources from sparsely
populated skymaps produced statistically convincing evidence for
small-scale clustering or correlations with any class of objects, nor
did they find a statistically convincing excess from any individual
astrophysical object.  ``Statistically convincing'' should be
emphasized here, as there is actually no shortage of claims for
clustering, correlation with classes of astrophysical objects, and 
excesses from plausible sources.  
Among the ``signals'' that have repeatedly been reported in
recent years are (1) an excess of cosmic rays with energies around
$10^{18}$\,eV near the galactic center~\cite{hayashida:1999}, 
(2) evidence for clustering of arrival directions of cosmic rays with 
energies above $4\times 10^{19}$\,eV on small angular scales ($\simeq$ 
degrees)~\cite{agasa:1996,agasa:1999,agasa:2000,agasa:2001,tinyakov:2001,
agasa:2003}, and (3) significant correlations with objects of the BL Lacertae 
class of AGN~\cite{tinyakov:2001b,tinyakov:2002,gorbunov:2002}.  
Most of the analyses leading to these claims
were based on arrival directions recorded by AGASA, though they are not
necessarily claims by the AGASA collaboration: the list of AGASA events
above $4\times 10^{19}$\,eV is one of the published and openly accessible
lists of cosmic ray events currently available.  Whenever these claims 
have been subjected to tests with statistically independent data, they 
have failed.  Much earlier there were suggestions of significant first 
and second harmonics in right ascension, suggestive of large-scale 
celestial anisotropy~\cite{fichtel:1986}, and
evidence for a flux from the direction of Cygnus X-3 was reported 
from two experiments ~\cite{cassiday:1989,teshima:1990}.  Recent 
experiments with much greater exposure and superior resolution have 
not confirmed any early indications of anisotropy.

We will now summarize and discuss several of the recent claims for
significant deviations from isotropy and subsequent tests of them with
new, statistically independent data.  We also analyze what ``went
wrong'' in the original analysis and what lessons can be learned for
current and future searches for anisotropy and correlations with
astrophysical sources.

\subsection{Galactic Center}

One of the most promising reports of a possible discovery of a
cosmic ray source is the announcement by the AGASA collaboration of a
$4\,\sigma$ excess of cosmic rays from the region of the galactic
center~\cite{hayashida:1999}.  The excess shows up in a very narrow
energy band, between 0.8\,EeV and 3.2\,EeV, while all neighboring
energy bins show no excess.  AGASA is located in the northern
hemisphere and the galactic center at declination
$\delta=-28.9^{\circ}$ is not in its field of view, but lies just
outside.  The excess was found by integrating event densities over a
$20^{\circ}$ radius at each sky location.  The significance of
$4\,\sigma$ should not be confused with a valid chance probability,
since the energy band and the integration radius are {\it a posteriori}, 
chosen to optimize the signal after analyzing the data.  No source 
flux is given in~\cite{hayashida:1999}.

Interest in this possible source, however, remained high for two
reasons.  First, in an independent analysis of archival data taken
between 1968 and 1979 by the SUGAR array in
Australia~\cite{bellido:2001}, an excess near the galactic center was
detected at the $2.9\,\sigma$ level.  The location of the excess did
not coincide with AGASA's location of maximum excess, but given the
large integration circle of AGASA's analysis and the angular
resolution of SUGAR, both excesses could stem from the same source.
Interestingly, the SUGAR excess is consistent with a point source.
Given the presence of galactic magnetic fields, this implies neutral
particles like neutrons and photons.  However, since SUGAR measures the 
secondary muons of the air shower underground, photon primaries with 
their small muon content are unlikely candidates.

The second reason is related to theoretical expectations.  The
galactic center region is a natural site for cosmic ray acceleration.
It contains a supermassive black hole, hosts a dense cluster of stars,
stellar remnants, and the supernova remnant Sgr A East.  In a striking
coincidence, the decay length of neutrons of energy $10^{18}$\,eV is
roughly 8.5\,kpc, the distance between us and the galactic center.
Neutron primaries would therefore offer an explanation for why the
signal only shows up in such a narrow energy band: neutrons below
$10^{18}$\,eV decay before reaching the Earth, and at energies higher
than several $10^{18}$\,eV, the source runs out of steam.
Furthermore, if the TeV gamma radiation recently observed by the
Cangaroo II~\cite{tsuchiya:2004}, H.E.S.S.~\cite{aharonian:2004}, and
Whipple~\cite{kosack:2004} air Cherenkov telescopes is produced
through photo-meson production with ambient photon fields, a flux of
$10^{18}$\,eV neutrons is
expected~\cite{aharonian:2004b,bossa:2003,grasso:2005,biermann:2004,medina:2004}.

With its southern location, the Pierre Auger Observatory is in an
excellent position to verify or rule out an excess of cosmic rays from
the galactic center.  A search for an excess from the galactic center
using the first 2.3 years of data~\cite{abraham:2007b} was one of the
first publications by the Pierre Auger experiment.  With a data set
already larger than AGASA's data set by a factor of 3 and larger than
the SUGAR data set by a factor of 10, the excess was not confirmed.
The data were used to set a source flux upper limit that excludes
several theoretical models of neutron
production~\cite{aharonian:2004b,bossa:2003} and the source flux given
by SUGAR~\cite{bellido:2001}.

\subsection{Small-Scale Clustering}

One way to try to identify the sources of ultrahigh energy cosmic rays
is to search for deviations from isotropy in the data set itself
rather than establishing matches with possible sources, i.e. to look
for an ``autocorrelation.''  As an example, if cosmic ray arrival
directions tend to ``cluster'' by showing more multiplets (doublets,
triplets,...)  of events than can be expected in a random isotropic
data set of the same size and exposure, it could indicate that cosmic
rays originate in nearby, discrete sources.  Attempts to find
small-scale clustering of ultrahigh energy cosmic rays above
$4\times 10^{19}$\,eV have repeatedly reported positive results
~\cite{agasa:1996,agasa:1999,agasa:2000,agasa:2001,tinyakov:2001,agasa:2003}.

\begin{figure}[t]
\centerline{\includegraphics[width=0.75\textwidth]{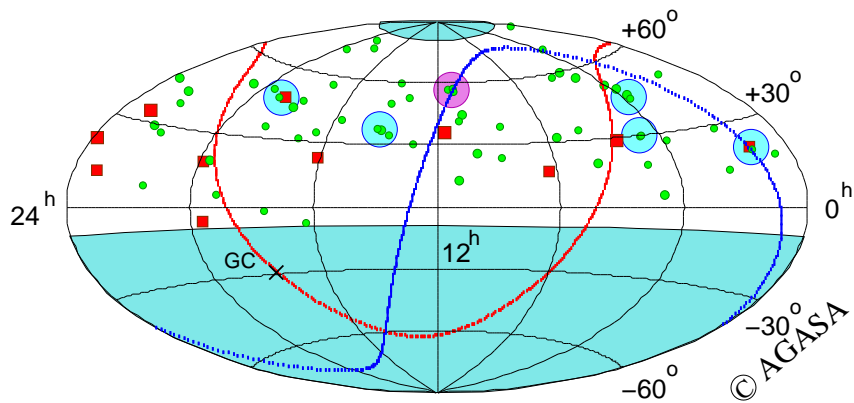}}
\caption{Skymap (in equatorial coordinates) of arrival directions of cosmic 
rays with energies above $4\times 10^{19}$\,eV recorded by the AGASA
experiment between 1990 and 2002.  Events with energies above 
$10^{20}$\,eV are shown in red.  Clusters are indicated as circles (blue for
doulets, pink for the triplet).  The galactic plane is indicated by the
red line, the supergalactic plane by the blue line.  
Taken from~\cite{agasa:web}.}
\label{fig:agasa}
\end{figure}

Searches for small-scale clustering usually involve the calculation of
the two-point-correlation function: for a given set of cosmic ray
events with energies above some threshold $E_{th}$, the number of
pairs $n_{p}$ separated by an angular distance less than $\theta$ is
counted and compared with the number expected in an isotropic
distribution of arrival directions.  This is used to calculate the
chance probability $P(E_{th},\theta)$ of finding $n_{p}$ or more pairs
in a random data set just by chance.  As an example, the published
AGASA data set of 57 events with energies above $4\times 10^{19}$\,eV
recorded until May 2000
contains 4 doublets and 1 triplet, where a doublet is defined as two
events with angular separation of less than $2.5^{\circ}$.  As
straightforward as this counting might sound, there is considerable
disagreement over the statistical significance of this potential
signal.  In fact, the literature gives chance probabilities between
$10^{-4}$ and 0.2.  Fig.\,\ref{fig:agasa} shows an updated skymap
with data taken until 2002, with doublets and triplets indicated.

The problem arises from the way the chance probability of the signal
is evaluated.  Several important parameters of the analysis are not
determined {\it a priori}, among them the maximum angular separation
of two events that defines a cluster, and the minimum energy of the
showers used in the analysis.  The optimal angular separation that
defines a cluster is unknown as magnetic smearing might cause
clustering to occur at angular scales somewhat larger than
the angular resolution of the instrument.  Unknown magnetic
deflections also make the choice for the energy threshold difficult,
because a higher energy threshold may reduce deflections of charged
cosmic ray primaries by magnetic fields, but it also weakens the
statistical power of the data set.  

Many authors have approached the problem by scanning the parameter
space for angular separations, energy thresholds and - in the case of
correlation searches - source catalogs that maximize a correlation.
This is a legitimate approach, but it does not yield the significance
of the potential signal.  Once a potential clustering signal is found
after scanning over one or more parameters, the probability that the
null hypothesis (``the sky is isotropic'') is rejected needs to be
evaluated with a statistically independent data set where the
parameters that were found {\it a posteriori} to maximize the
``signal'' in the original data set are now treated as {\it a priori}.
This is tough if the data set is small to begin with, and it is
tempting to abandon this rigorous procedure in favor of more or less
complete estimates of ``penalty factors'' resulting from the scan over
the parameter space.  

In the case of searches for small-scale clustering, this scan is
usually performed over some range of separation angles $\theta$ and
energy thresholds $E_{min}$ simultaneously.  In the first step, the
scan determines which energy threshold and angular separation
maximized the clustering signal.  Once found, the statistical
significance is then evaluated by performing identical scans over many
simulated isotropic data sets with the same exposure, and counting the
sets that have a stronger signal somewhere in the parameter space.

Such scans have been performed for the AGASA data
set~\cite{finley:2004}, the HiRes stereo data set~\cite{abbasi:2004},
and the Auger data set~\cite{mollerach:2007}.  For the HiRes data set
with 271 events above $10^{19}$\,eV, no significant clustering signal
is found for any separation angle between $0^{\circ}$ and $5^{\circ}$
and any energy threshold above $10^{19}$\,eV.  The same method gives a
chance probability of $3\times 10^{-3}$ for the strongest AGASA
clustering signal, above $4.9\times 10^{19}$\,eV with
$\theta=2.5^{\circ}$.  The Auger data (1 January 2004 - 15 March 2007)
were scanned above a minimum energy of $2\times 10^{19}$\,eV up to a
maximum $\theta$ of $30^{\circ}$.  The most significant signal, at
$\theta=7^{\circ}$ for $E>5.75\times 10^{19}$\,eV, has a chance
probability of $2\times 10^{-2}$.

Note that there are some hidden ``penalty'' factors that even this
method cannot account for.  Arrival direction data sets are analyzed
for possible anisotropies of many kinds.  Small-scale clustering is
just one example of a deviation from isotropy that an observer might
find interesting.  Other examples include events forming line-like
structures~\cite{abbasi:2007c}, or accumulating in particular regions
of the sky, or aligning with the galactic plane or the supergalactic
plane, or having a celestial dipole or quadrupole
dependence, or correlating with positions of particular accreting
neutron stars or black holes, or particular fast pulsars, or nearby
supernova remnants, or colliding galaxies, or gamma ray bursts, etc.
Astrophysical candidates can be targeted individually or ``stacked''
by targeting numerous ones collectively.  Correlation with some
catalog is an example of that.  There is no rigorous way to count the
types of anisotropy that are of potential interest.  Any finite data
set drawn from an isotropic distribution will have irregularities, and
thorough exploration is likely to result in some intriguing patterns
which occur very rarely in other data sets similarly sampled from
isotropy.  Without {\it a priori} specification of what to look for,
it is impossible to know how many tests were applied in recognizing an
intriguing pattern.  The danger is that a search without a pre-defined
goal may overestimate the significance of a feature that is discovered
by exploration.

The AGASA small-scale clustering signal, with a chance probability of
$3\times 10^{-3}$ as mentioned above is a good example for the perils
of estimating penalty factors.  As shown in~\cite{finley:2004}, the
chance probability for small-scale clustering actually increases to
almost 0.2 in an analysis that discards the data that were taken
before the first published claim of a clustering
signal~\cite{agasa:1996} and tests the claim using only data taken
afterwards.  Similar problems with published claims of significant
small-scale clustering have been pointed out by various
authors~\cite{finley:2004,watson:2001,evans:2003}.

At present, there is no confirmed evidence for significant clustering 
of cosmic ray arrival directions.  Clearly, any clustering of arrival 
directions is weaker than previously suggested.

\subsection{Maximum Likelihood Ratio}

While the two-point-correlation function has been widely used in
cosmic ray data analysis, it has some shortcomings.  It requires the
choice of an energy threshold and, by simply scanning over angular 
separations, it assumes that the angular resolution of the experiment
can be characterized by a single average value.  In reality, the angular 
resolution of cosmic ray detectors is not a constant, but rather depends
on a variety of parameters, most notably the energy and the position of 
the air shower relative to the detector.  Energy-dependent magnetic
deflections can also not be accounted for with a single rigid value for the
angular resolution.

The application of maximum likelihood techniques is a way to overcome 
some of these shortcomings.  As an example, a maximum likelihood ratio
test that makes full use of the instrument's point spread function has
been applied in searches for point-like sources in the combined HiRes 
and AGASA data set of cosmic rays with energies above $4\times 
10^{19}$\,eV~\cite{abbasi:2005}, and in the search for correlation of 
arrival directions with objects of the BL Lacertae class of 
AGN~\cite{abbasi:2006}.  The method can easily be extended, for example 
to allow for different source strengths in correlation 
studies~\cite{jansson:2007}.

In the search for cosmic ray point sources, the likelihood ratio
${\mathcal L}=P(\mathrm{Data}|\mathrm{H_1})/
P(\mathrm{Data}|\mathrm{H_0})$ is calculated for a given position
$\vec{x}$ on the sky.  Here, $P(\mathrm{Data}|\mathrm{H_1})$ is the
likelihood for the source hypothesis (``a source at position $\vec{x}$
contributes $n_{s}$ source events in addition to the expected
background'') and $P(\mathrm{Data}|\mathrm{H_0})$ is the likelihood of
the null hypothesis (``the event density at $\vec{x}$ is due to
background'').  One can now maximize the likelihood ratio as a
function of $n_s$ to obtain the best estimate for the number of events
contributed by a source at $\vec{x}$.  This estimate can then be
calculated for a dense grid of points on the sky covering the full
range of equatorial coordinates $\alpha$ (right ascension) and $\delta$ 
(declination) accessible to the experiments.  The parameters 
$\alpha$, $\delta$, and $n_{s}$ which maximize the likelihood ratio will 
therefore give us the best estimate for the position of a source and the 
number of events it contributes.

The calculation of ${\mathcal L}$ utilizes the arrival direction
accuracy for each event and the background
probability for each position on the sky.  It therefore uses all
available information for each event.  Additional factors, for example
the deflection of charged cosmic ray primaries in magnetic fields, can
be included even if the magnetic fields that cause the deflection are poorly
known.  This can be done by treating the magnetic field strength as a
nuisance parameter that is removed by marginalizing over all possible magnetic
fields.  The marginalized likelihood ratio automatically accounts for the
statistical penalty from the consideration of many magnetic field models.
Marginalized likelihood ratios are a common tool in many areas of
parameter estimation, see for example~\cite{lewis:2002}.

The analysis of the combined HiRes and AGASA data above $4\times
10^{19}$\,eV does not reveal evidence for a statistically significant
point source of cosmic rays.  The strongest ``hot spot'' is found to
have a chance probability of 28\,\% to appear in a random isotropic
data set~\cite{abbasi:2005}.

In searches for correlations with known astrophysical objects, there
are additional choices to be made in the selection of the classes of
possible sources.  Besides the choice of astrophysical objects, there
are, for example, choices for the minimum magnitude or maximum
distance of sources that are deemed likely candidates for correlations
with cosmic rays.  The search for correlations of ultrahigh energy cosmic
rays with objects of the BL Lacertae class of AGN is an example of a 
search where the selection criteria were changed with each new analysis.
Sources were first selected based on redshift ($z>0.1$ or unknown), 
optical magnitude ($m<18$), and 6\,cm radio flux 
($F_6>0.17\,\mathrm{Jy}$)~\cite{tinyakov:2001b}, then on optical magnitude 
alone ($m<18$)~\cite{tinyakov:2002}, then, in~\cite{gorbunov:2002}, 
on their possible association with gamma ray sources from the 
$3^{\mathrm{rd}}$ EGRET Catalog~\cite{hartman:1999}.  Each search
resulted in significant claims, mainly based on data recorded by AGASA 
and the Yakutsk experiment, applying various energy cuts.
Several authors have pointed 
out the dangers of this approach~\cite{evans:2003,stern:2005}, and 
statistically independent data sets subsequently did not confirm the 
correlations~\cite{torres:2003,abbasi:2006}.

A particular correlation first claimed in~\cite{gorbunov:2004} based
on the HiRes stereoscopic data set is not ruled out at this point.  In
this analysis, the 157 confirmed BL Lac objects with optical magnitude
$m<18$ from the $10^{\mathrm{th}}$ Catalog of Quasars and Active Nuclei 
by V\'eron-Cetty and V\'eron~\cite{veron:2001}.
were found to show correlations with the arrival directions of the 271
cosmic rays with energy $E>10^{19}$\,eV taken between December 1999
and January 2004.  The correlation is analyzed in detail
in~\cite{abbasi:2006}, where the authors also point out that
statistically independent data is necessary to estimate the
significance of the correlation.  Independent stereoscopic data taken
after January 2004 did not confirm the correlation, but since the
independent data set is smaller than the original and the HiRes
experiment was discontinued in 2006, the correlation claim is not
ruled out at a satisfying confidence level.

Data from the Pierre Auger Observatory was used to search for an
equivalent signal in the southern hemisphere, and no correlation was
found~\cite{harari:2007}, but it has been pointed out that the smaller
number of confirmed BL Lac objects in the southern sky requires a
larger data set to confirm or disprove the correlation
signal~\cite{gorbunov:2006}.

\subsection{Previous Claims: What could have gone wrong?}

Cosmic ray anisotropy detections have been made by researchers in
good faith, believing that they have uncovered essential clues to the
origins of the mysterious high energy charged particles.  Scientists
are highly motivated to discover the sources, and they feel an
obligation to report any clues that they find in the data.  A person
who toils for years to build a more powerful detector than any
predecessor naturally does so in hopes of finding new clues.  There
is a strong psychological preference for a positive result as opposed 
to a null result.

After exploring a data set thoroughly, some intriguing pattern is
likely to be found.  There is a long list of potential objects and
also many different classes of astrophysical objects that
one might consider as sources.  There are many possible signatures for
autocorrelation.  There are many large scale patterns to consider.
For each kind of potential anisotropy, a thorough search would try all
possible cuts on energy.  For many of them, angular separation cuts
would be varied.  An unrestricted exploration of a data set entails a
huge number of trials, and the effective number of trials cannot be
estimated.  When you look at a skymap, you will see a pattern that is
there.  But how many other patterns of equal or greater interest might
you have found if they had been there?  There is no way to count them.

Motivated by the desire to find new clues, it is natural to cite
reasons why a discovered pattern is particularly interesting.  It may
seem that the observed pattern should have been one of the first
things for which to look.  In this {\it a posteriori} analysis, it is then
natural to argue that the discovered pattern would have been found
with very few trials since any sensible scientist would have looked
for it with high priority.  Similarly, one may find reasons why the
energy cuts and angular cuts that maximize the ``signal'' are
physically reasonable.  They can seem like natural choices to
make, so perhaps only a small statistical penalty (or no penalty at
all) need be assessed for discovering the signal at its maximum
significance.

Confidence that a signal is genuine can lead to overlooking other
types of trials.  In studying the behavior of the signal, one might
find that it is enhanced by an alternative event reconstruction
technique, by different reconstruction quality cuts, by different
zenith angle cuts, by discarding (or including) certain epochs when
the detector was irregular in some respect, etc.  For optical
detectors, maybe weather cuts or atmospheric quality cuts can be
adjusted to enhance the signal.  A stronger signal may be regarded as
reason enough to accept such optimization of quality cuts as sensible,
if one is certain that the signal is genuine.  The quality cuts might
be reported without mentioning that they were tuned to enhance the
signal.  Strong confidence that an apparent signal is genuine might
come, for example, from a published model that calls for the
discovered anisotropy or from corroborating results reported by
another experiment (bandwagon effect).

The estimated statistical significance of anisotropy reported in {\it
a posteriori} analyses should be viewed with skepticism.  Confirmation
of the anisotropy by clearly specified procedures applied to
independent data is essential.

\subsection{Results from the Pierre Auger Observatory}
\label{sec:auger_agn}

Recently, the Pierre Auger Observatory published first results of
anisotropy studies based on data taken between January 2004 and August 
2007~\cite{abraham:2007c,abraham:2007d}.  The analysis shows
evidence for a correlation between the arrival directions of cosmic 
rays with energies above $6\times 10^{19}$\,eV and Active Galactic 
Nuclei with distances less than 75\,Mpc taken from the 
$12^{\mathrm{th}}$ Catalog of Quasars and Active Nuclei by
V\'eron-Cetty and V\'eron~\cite{veron:2006}.  The analysis finds 
that the hypothesis of an isotropic arrival direction distribution
of the highest energy cosmic rays is rejected at the 99\,\% confidence 
level.

To avoid the pitfalls of earlier analyses, the correlation search was 
performed in two steps.  An exploratory analysis first identified the 
angular separation, energy threshold, and maximum AGN redshift that 
maximize the correlation signal.  The data used to fine-tune the cut 
parameters were then discarded, and the correlation was tested on new 
independent data for which all cuts were considered {\it a priori}.  
The test was performed as a sequential analysis where the correlation
signal is evaluated after each new event, and the test stopped after 
the null hypothesis (no correlation) was rejected at least at the 99\,\%
confidence level.

\begin{figure}[t]
\centerline{\includegraphics[width=0.85\textwidth]{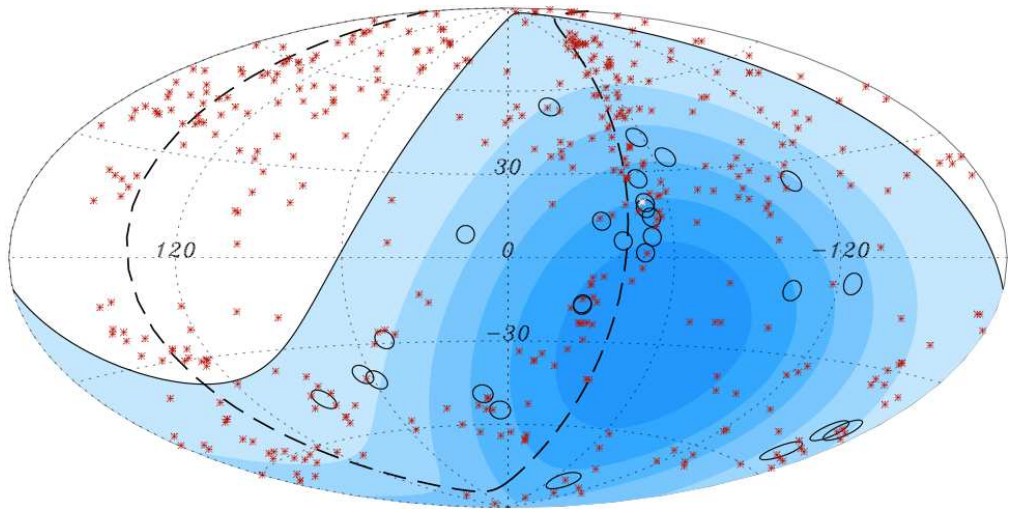}}
\caption{Skymap (in galactic coordinates) of arrival directions of cosmic 
rays with energies above $5.7\times 10^{19}$\,eV recorded by the Pierre 
Auger Observatory, and positions of Active Galactic Nuclei with redshift 
$z\leq 0.017$.  The cosmic ray events are shown as circles of radius 
$3.2^{\circ}$ radius centered on the arrival direction, the AGN are shown 
as red asterisks.  The blue shading indicates the relative exposure of the 
Observatory, with darker color indicating larger exposure.  The dashed line
indicates the supergalactic plane, and the position of Centarus A is marked 
by a white star.  Taken from~\cite{abraham:2007d}.}
\label{fig:auger}
\end{figure}

At present, based on data taken between 1 January 2004 and 31 August 2007,
the correlation signal has maximum strength for cosmic rays with
energies greater than $5.7\times 10^{19}$\,eV and AGN with redshift
$z\leq 0.017$, with a maximum angular distance that defines a correlation
of $3.2^{\circ}$.  For these parameters, 20 out of 27 events correlate
with the positions of AGN.  As the data set is still small, these
parameters merely give a first estimate of the relevant angular scale
and energy threshold.  Fig.\,\ref{fig:auger} shows a skymap of the
cosmic ray arrival directions and the positions of AGN with $z\leq 0.017$,
together with the relative exposure of the Auger Observatory.

The Auger correlation signal shows that the cosmic ray sky
at the highest energies is not isotropic and that charged particle 
astronomy at these energies is indeed possible.  While a firm
interpretation of the signal has to await additional data, the signal
already shows some remarkable features.  The energy threshold that
maximizes the correlation signal coincides with the energy of the GZK 
suppression, defined here as the energy where the flux has dropped to 
50\,\% compared to a simple power law extrapolation of the spectrum.
This further supports the interpretation that the steepening of the
spectrum at $6\times 10^{19}$\,eV is indeed due to the GZK effect rather
than an intrinsic feature of the source spectra.  The angular scale of the 
correlation signal of $3.2^{\circ}$ indicates that deflections in magnetic 
fields are small and the cosmic ray primaries are not predominantly heavy 
nuclei.  At this point, however, it is unclear whether AGN are the sources 
or act merely as tracers of the true sources.  The distribution of matter
in the nearby Universe is not isotropic, and the distribution of the 
nearby AGN correlates with the matter distribution.  Furthermore, the AGN 
catalog that forms the basis of this analysis is known to be incomplete,
in particular near the galactic plane.  While not relevant for the statistical 
evaluation of the correlation signal, the incompleteness is an obstacle on 
the way towards an interpretation of the signal.

While the correlation signal is a first indication that most cosmic rays 
with energies above $6\times 10^{19}$\,eV are protons from nearby AGN or 
sources with a similar spatial distribution, many questions remain open.
With the small number of events, distinctive properties of the subset of
AGN that correlate with cosmic rays cannot yet be identified.  While two 
events come from within $3^{\circ}$ of Centaurus A (at 3.4\,Mpc distance 
one of the closest AGN), there are more distant AGN that lie in almost that 
same direction.  Moreover, no event has been observed from the direction of 
the Virgo supercluster, which contains the giant elliptical galaxy M\,87
at a distance of only 18\,Mpc.

The Pierre Auger Observatory is only at the very beginning of its life.  
The data used for the current analysis corresponds to roughly 1.2 years of 
operation of the complete detector.  In little more than a year, the exposure 
will already have doubled.  With more data, an unambiguous identification of 
the sources might be possible, and individual sources might start to emerge.

\section{Ultrahigh Energy Gamma Rays and Neutrinos}
\label{sec:neutrino}

In the previous chapter, we summarized the current status of astronomy
with cosmic rays at ultrahigh energies.  Cosmic rays are not the
only messenger particles that potentially arrive from cosmic particle
accelerators.  If (charged) cosmic rays are accelerated in astrophysical 
objects like supernova remnants or AGN, high energy gamma rays and 
neutrinos are secondary products of this acceleration.  When high energy 
protons hit ambient gas and radiation in the vicinity of the acceleration 
sites, the production and subsequent decay of charged and neutral pions
produces gamma rays and neutrinos.  Gamma rays are also produced
by bremsstrahlung and synchrotron losses, and inverse Compton
scattering can upscatter photons to higher energies.

We will now discuss the possibility of astronomy with gamma rays 
and neutrinos at ultrahigh energies.

\subsection{Gamma Rays}
\label{subsec:gamma}

Galactic and extragalactic sources of gamma rays with energies up
to tens of TeV have been identified with high statistical significance
by air Cherenkov telescopes for several years now.  Lately, several
extended sources of gamma rays have been identified by the Milagro
gamma ray allsky survey~\cite{atkins:2005}.  The gamma rays observed by
these experiments are the highest energy particles seen that we can
associate with discrete astrophysical sources.

At energies above several tens of TeV, the Universe is no longer
transparent to gamma rays as they interact with extragalactic
background photons.  At 20\,TeV, the mean free path of gamma rays
on the infrared background is only about 100\,Mpc, and above 100\,TeV,
interaction with the microwave background leads to an even more
dramatic drop in the mean free path, down to about 10\,kpc near
$10^{15}$\,eV.  At much higher energies, above $10^{19}$\,eV, the 
isotropic radio photons become the dominant background targets, 
but the attenuation length remains well below 100\,Mpc, with
some uncertainty due to our poor knowledge of the radio background.

The prospects for extragalactic gamma ray astronomy at energies
above several tens of TeV are therefore dim.  Nevertheless, there are
many scenarios for cosmic ray production that predict a substantial
fraction of gamma rays in the cosmic ray flux above $10^{19}$\,eV.
Top-down models are an example.
Partly created to address the problem that particle acceleration to 
energies above $10^{20}$\,eV by electromagnetic processes is difficult
in known astrophysical sources, these models also offer an explanation 
for the observed isotropy of arrival directions and the extension of the 
spectrum beyond GZK energies as seen by AGASA.  Rather than assume a 
``bottom-up'' acceleration, top-down models assume that ultrahigh energy 
cosmic rays are the decay products of supermassive particles, so-called 
$X$ particles, that can have masses up to $10^{24}$\,eV.  Examples are 
magnetic monopoles and other topological defects that might have been 
produced in the early stages of the Universe; see~\cite{sigl:2000} 
for a review of these models.   $X$ particles decay into quark-antiquark
pairs, producing two jets with about 95\,\% pions and 5\,\% baryons.
The majority of ultrahigh energy particles detected on Earth should 
consequently be photons from pion decay.

Another model that predicts photon primaries is the Z-burst 
model~\cite{weiler:1982}, which is an attempt to explain the ultrahigh 
energy cosmic ray flux and its features without the need for acceleration.  
The model is based on the idea that ultrahigh energy cosmic ray neutrinos 
interact with the relic neutrino background, generating $Z$ bosons that 
immediately decay with photons of ultrahigh energy as the main decay product. 
The theoretical motivation for some of these models has waned with the
accumulating evidence for the GZK cutoff. 

Up to recently, upper limits on the fraction of photons in the cosmic ray
flux did not impose serious constraints on top-down models or the Z-burst
scenario.  Data from the Haverah Park experiment were used to set
upper limits on the photon fraction of 48\,\% above 10\,EeV and 50\,\% above 
40\,EeV~\cite{ave:2000,ave:2002}, and AGASA data were used to set upper 
limits of 28\,\% above 10\,EeV, 67\,\% above\,32\,EeV~\cite{shinozaki:2002}, 
and 67\,\% above 125\,EeV~\cite{risse:2005} (all limits at 95\,\% c.l.).

The situation has changed with the influx of data from the Pierre Auger 
Observatory.  New upper limits have now been reported from fluorescence 
detector and surface detector data~\cite{abraham:2007,abraham:2007e} based 
on measurements of the height of the shower maximum.  On average, the 
depth of shower maximum for 10\,EeV photons is $1000\,\mathrm{g\,cm^{-2}}$ 
as compared to $800\,\mathrm{g\,cm^{-2}}$ for protons of the same energy.  
At energies above about 10\,EeV, the Landau-Pomeranchuk-Migdal (LPM)
effect~\cite{landau:1953,migdal:1956}, 
a suppression of the Bethe-Heitler cross section, further delays the shower
development.  In addition, above about 50\,EeV, the primary photon can
interact with the geomagnetic field and convert to $e^{+}e^{-}$-pairs
which subsequently create more photons via bremsstrahlung, initiating
a pre-shower of particles above the atmosphere.  What ultimately
enters the atmosphere is not a single photon, but several low energy
particles.  This effect, which strongly depends on the energy and
incoming direction of the primary gamma ray relative to the local
geomagnetic field, can counteract the delay of the shower development
due to the LPM effect, and both effects need to be taken into account
when simulating photon primaries.

The air fluorescence detectors of the Pierre Auger Observatory
determine the depth of shower maximum directly.  For the surface
detector data, several observables indirectly indicate the height of
the shower maximum, and the data taken in hybrid mode allow for a
careful cross-check of both direct and indirect techniques.

\begin{figure}[t]
\centerline{\includegraphics[width=0.7\textwidth]{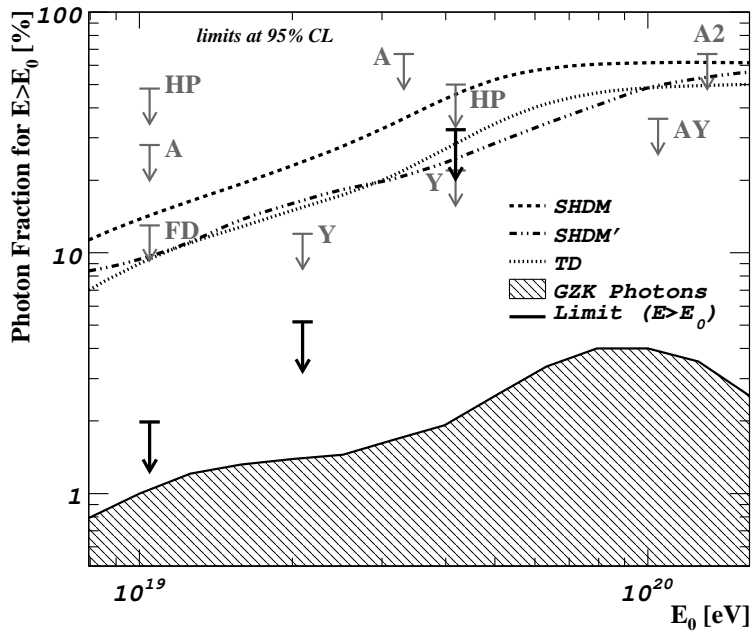}}
\caption{95\,\% confidence limit upper limits on the fraction of photons 
in the integral cosmic ray flux, as a function of energy, from data taken with
the surface detector array of the Pierre Auger Observatory~\cite{abraham:2007e}
(black arrows).  For comparison, the plot also shows previous experimental
limits, from Haverah Park (HP)~\cite{ave:2000}, 
AGASA (A1, A2)~\cite{shinozaki:2002,risse:2005}, 
AGASA-Yakutsk (AY)~\cite{rubtsov:2006}, 
Yakutsk (Y)~\cite{glushkov:2007},
and Auger hybrid data (FD)~\cite{abraham:2007}.
Predictions from top-down models and predictions
of the GZK photon flux are also shown.  Taken from~\cite{abraham:2007e}.}
\label{fig:photon}
\end{figure}

With hybrid data, an upper limit of 16\,\% for the photon fraction in the cosmic ray
flux above $10^{19}$\,eV was obtained at 95\,\% confidence level~\cite{abraham:2007}.
The larger data set taken with the surface detector array alone yielded upper limits 
of 2.0\,\%, 5.1\,\%, and 31\,\% at $10^{19}$\,eV, $2\times 10^{19}$\,eV, and 
$4\times 10^{19}$\,eV, respectively~\cite{abraham:2007e}.  Fig.\,\ref{fig:photon}
shows these limits along with limits from previous experiments, predictions
from top-down models, and predictions for the flux of photons produced through
photoproduction of pions in the interaction of cosmic rays with the
microwave background (``GZK photons'').

\subsection{Cosmogenic Neutrinos}

Neutrinos are not only produced in the cosmic ray sources themselves,
but also during cosmic ray propagation.  Cosmic rays lose energy by
photoproduction of mesons on extragalactic photons, and
neutrinos are an end product of meson decay.  The same interaction
that is responsible for the GZK cutoff in the energy spectrum of
ultrahigh energy cosmic rays therefore inevitably produces a diffuse
flux of neutrinos at ultrahigh
energies~\cite{berezynski:1969,stecker:1973}.  Detecting these ``GZK
neutrinos'' is the goal of several ongoing and planned experiments.

It is important to note that the neutrino cross section grows with energy, and 
at ultrahigh energies, neutrinos can no longer penetrate the Earth.  Detecting
upward-going muons, a technique applied at lower neutrino energies, therefore
no longer works for ultrahigh energy neutrinos at $10^{18}$\,eV.  The detection 
of a neutrino at such high energies is only possible if the neutrino interacts in 
close proximity to the detector producing an electromagnetic cascade.  The
challenge is to find ways to pick up this cascade with a detection technique 
that allows to probe very large volumes without the need to densely instrument
the volume.

Experiments have recently started to use a detection technique that was suggested 
as early as 1962 by G. Askaryan~\cite{askaryan:1962,askaryan:1965}.  The electromagnetic 
cascade that develops after the neutrino interaction is not entirely symmetric in charge,
as electrons and positrons have different scattering properties and interactions.  
The net excess of electrons over positrons means that the shower as a whole resembles
a negative charge moving at nearly the speed of light, producing Cherenkov radiation
as a consequence.  At wavelengths larger than the extent of the showers 
themselves, the emission is coherent; for a typical solid medium, the emission is coherent 
at radio wavelengths.  It is crucial that the medium in which the radio emission is 
produced does not attenuate the radio signal appreciably.  Ice is a suitable material,
and several ongoing and future experiments use the Antarctic ice as the detector medium.

\subsection{Detection of GZK Neutrinos}

A dedicated instrument to study the cosmogenic neutrino flux is the Antarctic
Impulsive Transient Antenna (ANITA) experiment~\cite{miocinovic:2004}.  ANITA 
detects neutrinos interacting in the Antarctic ice sheet by their coherent radio
signal via the Askaryan effect described above.
With an attenuation length of more than a kilometer, the ice is practically 
transparent for radio waves, and the radio pulses from the cascade can be picked 
up by antennas probing the ice surface from a balloon at an altitude of about 
37\,km.  With this technique, about $2\times 10^{6}\, \mathrm{km}^{3}$ of volume 
can be probed instantaneously, basically the entire visible area of the ice 
sheet up to a depth of about one attenuation length.  The existence of the Askaryan 
effect in ice was confirmed in laboratory experiments before the first 
launch~\cite{gorham:2006}.

While the full instrument was first launched in December 2006, a smaller prototype,
ANITA-lite, completed a successful 18.4 day flight in 2004.  The upper limit at 
90\,\% confidence level on the flux $F$ of cosmogenic neutrinos, $E_{\nu}^2\cdot 
F\leq 1.6\times 10^{-6}\,\mathrm{GeV}\, \mathrm{cm}^{-2}\,\mathrm{s}^{-1}\,
\mathrm{sr}^{-1}$ for energies $10^{18.5}\,\mathrm{eV}\leq E \leq 10^{23.4}\,
\mathrm{eV}$, based on this data set~\cite{barwick:2006a} excludes several 
``exotic'' models of ultrahigh energy cosmic ray production, most notably the 
Z-burst model, and improves considerably on the limits achieved with earlier 
experiments like GLUE~\cite{gorham:2004} and FORTE~\cite{lehtinen:2004}.
With the full instrument, ANITA is expected to observe 5 to 50 GZK neutrinos for
50 days of flight time, bringing the sensitivity to the level where a detection of 
the cosmogenic neutrino flux is expected.

The observation time constraints imposed by the limited duration of balloon flights can be 
overcome by placing radio detectors in suitable configurations on the ground.  The recently 
proposed Antarctic Ross Iceshelf Antenna Neutrino Array (ARIANNA)~\cite{barwick:2006b} plans 
to exploit the unique features of the Ross Ice Shelf near the Antarctic coast to increase the
sensitivity for the detection of GZK neutrinos by an order of magnitude over ANITA.  In its 
current design, ARIANNA consists of $100\times 100$ antennas on a square grid with 300\,m 
spacing.  Buried just under the surface, the antennas can detect direct radio emission from 
(nearly horizontal) neutrinos interacting in the shelf, or radio emission from downward-going 
neutrinos that is reflected at the water-ice boundary below the shelf.  The thickness of the 
shelf is about 500\,m.  First estimates of the sensitivity of the instrument show that ARIANNA 
reaches down in energies to almost $10^{17}$\,eV, bridging the energy gap between IceCube and 
ANITA.

\subsection{Pierre Auger Observatory as a Neutrino Detector}

While primarily a cosmic ray detector, the Pierre Auger Observatory is
also sensitive to neutrinos at EeV energies and
above~\cite{bertou:2002,abraham:2007f,muniz:2007}, the energy range
relevant for the search for GZK neutrinos.  If a $\tau$-neutrino
interacts in the mountains or the ground near the Observatory, the
$\tau$ lepton is expected to travel tens of kilometers before decaying
and producing an air shower.  Because of the Earth's opacity to EeV
neutrinos, such $\tau$-induced showers are nearly horizontal and can
be detected readily above the Auger array.  The showers caused by
neutrino interactions differ markedly from very inclined showers caused
by cosmic ray nuclei.  A cosmic ray air shower at large zenith angle
reaches ground level at a very large slant depth.  The electromagnetic
component is extinguished and only the muonic component survives.  These
showers are referred to as ``old'' showers.  A ``young'' 
shower due to $\tau$ decay is largely electromagnetic.
The young electromagnetic shower produces signals in the tanks that
are of relatively long duration.  Since the signal in the surface
detector tanks is read out by flash-analog-to-digital-converters
(FADCs), narrow signals from cosmic ray showers and broad signals from
earth-skimming $\nu_{\tau}$ can be distinguished. 
The calculation of the detector acceptance for $\nu_{\tau}$ is tricky.
Only neutrinos from a small fraction of the sky, essentially nearly
horizontal showers, can trigger the detector, and the interaction in
the Earth has to occur at a distance from the detector that allows the 
$\tau$-decay air shower to develop above the array where it can be 
measured.  Computer simulations give an acceptance of several times
$10^{16}\,\mathrm{cm}^{2}\,\mathrm{s}\,\mathrm{sr}$ above
$10^{19}$\,eV, with large systematic uncertainties from the $\nu$
cross sections and the fact that the actual topography of the site is
only included to some approximation.  In the data set taken since
January 2004, no earth-skimming $\nu_{\tau}$-event is seen, and a
90\,\% confidence level upper limit of $E_{\nu}^2\cdot
dN_{\nu}/dE_{\nu} \leq 2.0\times 10^{-7}\, \mathrm{GeV}
\,\mathrm{cm}^{-2}\,\mathrm{s}^{-1}\, \mathrm{sr}^{-1}$ 
is set~\cite{abraham:2007f}.  It should be noted that this limit is based on 
the $\nu_{\tau}$ cross section from~\cite{anchordoqui:2006}.  Since no 
measurement exists at these energies, the systematic error on the cross 
section is large, and the limit does not account for possible large
deviations from the value in~\cite{anchordoqui:2006}.  The limit is still 
well above theoretical expectations for the GZK neutrino flux, but is 
expected to improve by an order of magnitude over the lifetime of the 
Observatory.

In addition to earth-skimming $\nu_{\tau}$, there is also some 
sensitivity to downward-going neutrinos of all flavors.  Again, 
at large zenith angles, these neutrino-induced showers can be 
distinguished from cosmic ray air showers by their electromagnetic 
component which appears as a broad signal in time in the FADCs.  
A search for down-going showers initiated by neutrinos and a study 
of the detector sensitivity for this component is 
ongoing~\cite{muniz:2007}.

\section{Full-sky observatories}
\label{sec:fullsky}

Cosmic ray astronomy must cover the entire sky.  High energy cosmic
ray experiments have historically focused on the energy spectrum and
the chemical composition, and most detectors have been built in the
northern hemisphere.  Those include Volcano Ranch, Haverah Park,
Yakutsk, Akeno and AGASA, Fly's Eye and HiRes.  Before the Auger
Observatory, SUGAR (in Australia) was the only array to study cosmic
rays at the highest energies in the southern hemisphere.  Now the
Auger Observatory site in Argentina has surpassed the cumulative
exposure of the northern detectors, and over the next few years it
will accumulate a data set which exceeds previous experiments by more
than an order of magnitude.  The southern sky has become over-observed
relative to the northern sky.

The Auger Observatory was designed to have sites in both hemispheres
in order to have nearly uniform coverage over the celestial sphere.
The Auger plan is still to construct a site in southeast Colorado.  If
it is substantially larger than the Argentina site, then the northern
exposure could catch up to the southern.

For cosmic rays above $10^{19}$\,eV it is possible to achieve a huge
aperture by looking down on the atmosphere with a fluorescence
detector satellite in space.  The idea was first proposed by Linsley
\cite{linsley:1979} and developed later by the OWL Collaboration
\cite{stecker:2004} and EUSO~\cite{catalano:2003}.  The concept is now 
being developed most strongly as JEM-EUSO~\cite{takizawa:2007}.  An orbiting 
satellite looks at showers in the northern and southern hemispheres equally, 
and it should achieve nearly uniform exposure to the celestial sphere.

Full-sky exposure is important for a variety of reasons:

\noindent $\bullet$ The cosmic ray source giving the greatest flux at 
Earth could be in any part of the sky.  It is obviously advantageous 
to have a cosmic ray observatory well positioned to measure particles 
arriving from the brightest source.

\noindent $\bullet$ If no sources are to be detected, it is valuable
to have uniform flux limits over the sky.

\noindent $\bullet$ If discrete sources are detectable, making an
unbiased catalog of them requires good exposure to the full sky.

\noindent $\bullet$ Whether or not discrete sources are detected, it
is important to measure whatever other anisotropy is present.  Mapping
the sky in various energy bins would provide invaluable constraints on
theories about the sources and propagation of the cosmic rays through
magnetic fields.

\noindent $\bullet$ Celestial scatter plots have intuitive meaning
only if the exposure is approximately uniform.  A steep exposure
gradient makes it impossible to assess density patterns of arrival
directions.  

\noindent $\bullet$ Similarly, mapping the smoothed density function
is unsatisfying if the density of arrivals is mostly tracking the
exposure function.  Plotting the statistical significance of density
excess is a way to look for discrete sources, albeit with greatly
different sensitivity in parts of the sky that differ greatly in
exposure.  Such a map does not help in evaluating large scale
anisotropy.

\noindent $\bullet$ Multipole moments (the coefficients in a spherical
harmonic expansion of anisotropy) are the natural way to characterize
a celestial anisotropy.  The celestial intensity function can be fully
specified by a table of coefficients up to the order ($l$) for which
any structure would be smaller than the detector's angular resolution.
Measuring the multipole moments does not require uniform exposure, but
it does require full-sky coverage (see \cite{sommers:2001} for details).
Each multipole is a $Y_{lm}$-weighted integral of the cosmic ray
intensity over the sphere of arrival directions.  If there is a patch
of the celestial sphere with undetermined intensity, then none of the
multipoles can be measured.  Assumptions must be made about the
behavior in the missing patch in order to make even an unbiased
estimate of any multipole.  With full-sky coverage, the anisotropy can
be measured unambiguously and characterized in a way that can be
readily used to test theories.

\noindent $\bullet$ The angular power spectrum is a
coordinate-independent measure of the amount of anisotropy on angular
scales of 1/$l$ radians.  The ``power'' $C_l$ is simply the average squared
multipole at order $l$.  An unambiguous measure of the angular
power spectrum requires exposure to the entire sphere of arrival
directions.

\noindent $\bullet$ For large-scale patterns, the dipole and
quadrupole moments may be the most important contributors.  It is
clearly not possible to measure these moments if a quarter of the sky
is not exposed at all.  In that case, observers are sometimes forced
to find the dipole which best fits the intensity measured over the
exposed part of the sky, {\it assuming the pattern to be a pure
dipole}.  If a quadrupole pattern is suspected (e.g. an excess from
galactic equatorial regions), then the data might be tested against
that hypothesis.  In fact, with a detector in only one hemisphere, it
can be difficult to distinguish a pure quadrupole from a pure dipole
\cite{wdowczyk:1984}.  Without full-sky coverage, the search for
large-scale anisotropy reduces essentially to hypothesis testing or to
fitting a prescribed functional form.  {\it Measuring} the celestial
anisotropy requires exposure to the full sky.  The full power of
spherical harmonic expansion of the celestial intensity function is
then available to characterize the anisotropy.

\section{Summary}
\label{sec:summary}


Cosmic ray astronomy is ultimately the study of individual cosmic ray
sources via the particles that come from them.  It should also be
possible to infer properties of the intervening magnetic fields.  The
best hope for studying discrete sources this way is at the highest
cosmic ray energies, even though the fluxes there are minuscule.  It
requires enormous exposure to acquire adequate statistics at extremely
high energy, but there are reasons why this kind of astronomy may be
plausible there.  First of all, one can take advantage of the GZK
effect to eliminate the isotropic background caused by sources
throughout the distant Universe.  The GZK distance limit restricts
arrival directions to foreground sources in our part of the Universe.
Secondly, protons near and above the GZK energy threshold have
sufficient magnetic rigidity to penetrate through the galactic
magnetic fields without being deflected more than a couple of degrees.
Although the magnitude and coherence of intergalactic fields are
poorly known, such high energy protons would probably not be deflected
more than a few degrees coming from any source within that GZK radius.
Together, these two reasons motivate a search for discrete sources at
energies above the GZK energy threshold.

There is no guarantee that cosmic rays are produced by discrete
sources.  It is conceivable that they originate diffusely throughout
the Universe.  Top-down scenarios of cosmic ray production by decay or
annihilation of super-heavy relic particles have been a prominent
possibility for diffuse cosmic ray production.  Those models are
constrained by the diffuse gamma ray intensity measured near 100 MeV.
Top-down models that explain all the high energy cosmic rays would
normally predict a higher-than-measured gamma ray background.  The
top-down models have recently been almost ruled out by new upper
limits on the photon component of cosmic rays near $10^{19}$\,eV.
Again, top-down models would normally exceed those limits.  In
principle, the density of high energy neutrinos can also be used to
test for top-down production, since pion decays produce neutrinos as
well as gamma rays.  In practice, however, neutrino upper limits are
much weaker than the gamma ray limits.  The new gamma ray limits and
evidence for the GZK suppression~\cite{abbasi:2007,yamamoto:2007} 
in the cosmic ray spectrum, are reasons to believe that cosmic rays 
are produced by a bottom-up acceleration process in special astrophysical 
sources.  This is not a guarantee, but a strong suggestion, that discrete
sources of cosmic rays are responsible for the particles that are
recorded at Earth.


The fact that no individual source has so far been detected means that 
cosmic ray astronomy is not easy and might require much greater exposure.  
It is easier to detect the existence of discrete sources statistically than
individually.  One way is through autocorrelation of arrival
directions.  Significant clustering of arrival directions might be the
first evidence for discrete sources.  Alternatively, a correlation of
arrival directions with a candidate set of astrophysical objects would
also demonstrate anisotropy indicative of discrete sources.  A correlation
of that type can occur without any clustering of arrival directions,
since it does not require having more than a single arrival direction
from any one of the candidate sources.  Data taken with the Pierre Auger
Observatory indeed show first evidence of cosmic ray anisotropy at 
ultrahigh energies.  The arrival directions of cosmic rays with energies
above $5.7\times 10^{19}$\,eV correlate with the positions of nearby
($z\leq0.017$) Active Galactic Nuclei.

The Pierre Auger Southern Observatory is nearing completion in
Argentina, and its cumulative exposure is growing rapidly and is
already large compared to that of any previous experiment.
Constructing a larger Auger Northern Observatory in Colorado could
soon provide exposure to the full sky.  That would allow the detection
of discrete sources with nearly uniform sensitivity over the full
celestial sphere.  Spherical harmonics could completely characterize
the large-scale anisotropy.  Although presently still only a hope, the
prospects for cosmic ray astronomy are bright.

\ack
The work has been supported in part by the National Science
Foundation under contract numbers PHY-0555317 and PHY-0500492.

\end{document}